\documentclass[letter]{aa}
\usepackage{times}
\usepackage{graphicx}
\usepackage{natbib}
\usepackage{amsmath}
\usepackage{lscape}
\def\msun{{\rm\,M_\odot}}
\def\kms{{\rm km\;s}^{-1}}

\begin{document}

\title{The dynamically hot stellar halo around NGC~3311: a small cluster-dominated
central galaxy\thanks{Based on observations collected at the European
    Organisation for Astronomical Research in the Southern Hemisphere,
    Chile, during the observing program 082.A-0255(A) on 2009 March 25.}}

   \subtitle{}

   \author{G. Ventimiglia \inst{1,2}
          \and O. Gerhard \inst{1}
          \and M. Arnaboldi \inst{2,3}
          \and L. Coccato \inst{1}
          }

   \offprints{G. Ventimiglia, e-mail: gventimi@mpe.mpg.de}

   \institute{Max-Plank-Institut f\"ur Extraterrestrische Physik,
     Giessenbachstra$\beta$e 1, D-85741 Garching bei M\"unchen, Germany.
     \and European Southern Observatory, Karl-Schwarzschild-Stra$\beta$e 2,
          85748 Garching bei M\"unchen, Germany.
     \and INAF, Osservatorio Astronomico di Pino Torinese, I-10025 Pino Torinese, Italy.
}

   \date{Received July 27, 2010;
     accepted September 10, 2010}


  \abstract
  {An important open question is the relation between intracluster
    light and the halos of central galaxies in galaxy clusters.}
  {Here we report results from an on going project with the aim to characterize the
    dynamical state in the core of the Hydra I (Abell 1060) cluster
    around NGC~3311.}
    {We analyze deep long-slit absorption line spectra reaching out to
      $\sim25$ kpc in the halo of NGC~3311.}
    {We find a very steep increase in the velocity dispersion profile
      from a central $\sigma_0=150\;\mbox{km s}^{-1}$ to $\sigma_{out}
      \simeq 450\;\mbox{km s}^{-1}$ at $R \simeq
        12\;\mbox{kpc}$. Farther out, to $\sim 25$ kpc, $\sigma$
      appears to be constant at this value, which is $\sim60\%$ of the
      velocity dispersion of the Hydra I galaxies. With its
      dynamically hot halo kinematics, NGC~3311 is unlike other normal
      early-type galaxies.}
    {These results and the large amount of dark matter inferred from
      X-rays around NGC~3311 suggest that the stellar halo of this
      galaxy is dominated by the central intracluster stars of the
      cluster, and that the transition from predominantly galaxy-bound
      stars to cluster stars occurs in the radial range 4 to 12
      kpc from the center of NGC~3311. We comment on the wide range of
      halo kinematics observed in cluster central galaxies, depending
      on the evolutionary state of their host clusters.}

   \keywords{galaxies:clusters:general -- galaxies:clusters:individual (Hydra~I) 
             -- galaxies:kinematics and dynamics -- galaxies:individual (NGC~3311)}

   \titlerunning{The dynamically hot stellar halo of NGC~3311}

   \authorrunning{Ventimiglia et al. }

   \maketitle

\section{Introduction}

An important open question is the physical and evolutionary relation between the
intracluster light (ICL) and the extended halo of the brightest cluster galaxies
(BCGs), whether they are truly independent components or whether the former is a
radial extension of the latter. Using a sample of 683 SDSS clusters
\cite{Zibetti05} found a surface brightness excess with respect to an inner
$\mbox{R}^{1/4}$ profile that characterizes the mean profile of the BCGs, but it
is not yet known whether this cD envelope is simply the central part of the
cluster's diffuse light component, or whether it is distinct from the ICL and
part of the host galaxy \citep{Gonzalez05}.

In the Southern hemisphere, the cD galaxy NGC~3311 and the giant elliptical
NGC~3309 dominate the central region of the Hydra~I cluster, an X-ray bright,
non-cooling flow, medium compact cluster with a velocity dispersion
$\sigma_{Hydra I}=784\;\mbox{km s}^{-1}$ \citep{Misgeld08}. The X-ray
observations show that the hot intracluster medium centered on NGC~3311 has a
fairly uniform distribution of temperature and metal abundance from a few kpc
out to a radius of $230$~kpc \citep{Tamura00,Yamasaki02,Hayakawa04,Hayakawa06}.
Given the overall regular X-ray emission and temperature profile, the Hydra~I
cluster is considered as a prototype of an evolved and dynamically relaxed
cluster; it is therefore a suitable candidate for a dynamical study of a relaxed
extended stellar halo around a BCG.

The primary goal of this work is to establish the dynamical state of the stellar
halo of NGC~3311. We use long-slit spectra to uncover the kinematics in the halo
region of NGC~3311 out to $\sim 25$ kpc from its center.  In
Sect.~\ref{sigma3311_cap} we present observations with FORS2 at VLT and the
GEMINI GMOS archive data, which we reanalyze. We describe the data reduction and
the kinematic measurements in Sect.~\ref{datared}. The newly measured halo
kinematics and their implications are discussed in Sect.~\ref{kinN3311}, and our
conclusions are summarized in Sect.~\ref{Conclu}.

We adopt a distance to NGC~3311 of $51\,\mbox{Mpc}$ (NED), equivalent to a
distance modulus of $33.54$ mag. Then $1''$ corresponds to $0.247\;\mbox{kpc}$.

\section{Observations and archive data}\label{sigma3311_cap}

\begin{figure}[hbt!] \centering
\includegraphics[width=7.0cm]{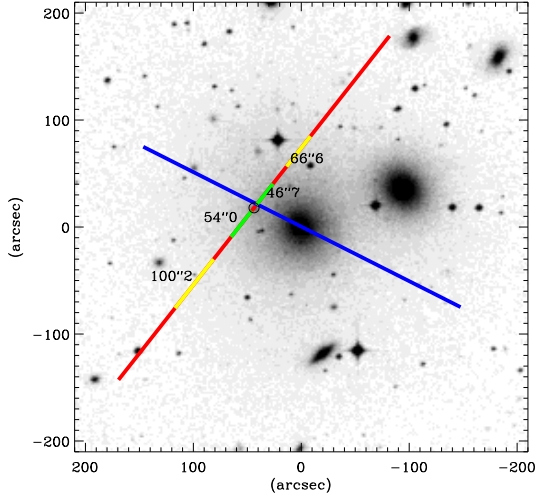}
\caption{Optical DSS $ 7' \times 7'$ image centered on NGC~3311 in the Hydra I
cluster. The relative positions of the GMOS slit ($5\farcm5$, blue line) and
FORS2 slit ($6\farcm8$, red line) are illustrated.  Green and yellow sections on
the FORS2 slit indicate regions where average spectra are extracted. The
adjacent numbers specify the radial distances of their light-weighted mean
positions from the center of NGC~3311. The center of the FORS2 slit coincides
with the position of the dwarf galaxy HCC~26 and is marked by a black circle.
North is up and East to the left.}
\label{geometry}
\end{figure}

{\it VLT FORS2 long slit spectra} - The long-slit spectra were obtained during
the nights of 2009 March 25-28, with FORS2 on VLT-UT1. The instrumental setup
had a long-slit $1\farcs6$ wide and $6\farcm8$ long, Grism 1400V+18, with
instrumental dispersion of 0.64 \AA\ pixel$^{-1}$ and spectral resolution
$\sigma=90$ $\mbox{km s}^{-1}$, and a spatial resolution along the slit of
$0\farcs252\;\mbox{pixel}^{-1}$. The seeing during observations ranged from
$0\farcs7$ to $1\farcs2$.  The wavelength coverage of the spectra is from 4655
\AA\ to 5965 \AA, including absorption lines from H$_{\beta}$, MgI
($\lambda\lambda 5167, 5173, 5184$ \AA) and Fe I ($\lambda\lambda 5270, 5328$
\AA). We obtained eight spectra of 1800 sec each, for a total exposure time of 4
hrs. In the FORS2 observations, the long slit is centered on the dwarf galaxy
HCC~26 at $\alpha=10\mbox{h}36\mbox{m}45.85\mbox{s}$ and $\delta=
-27\mbox{d}31\mbox{m}24.2\mbox{s}$ (J2000), with a position angle of
P.A.=$142^{\circ}$; HCC~26 is seen in projection onto the NGC~3311 halo. The
position of the FORS2 long slit is shown in Fig.~\ref{geometry}. Its center is
located at P.A.=$64^{\circ}$ with respect to NGC~3311, approximately along the
major axis of the galaxy.

{\it Gemini GMOS-South long slit spectra} - We use Gemini archive long-slit
spectra in the wavelength range from 3675 \AA\ to 6266 \AA\, observed with the
B600 grating, a dispersion of $0.914$ \AA\ pixel$^{-1}$, a spectral resolution
of $\sigma=135$ $\mbox{km s}^{-1}$, and a spatial scale of $0\farcs146$
pixel$^{-1}$; a detailed description of the instrumental setup is presented in
\cite{Loubser08}. The seeing was typically in the range from $0\farcs6$ to
$1\farcs2$.  We target the same absorption lines as for the FORS2 spectra, i.e.
H$_{\beta}$, MgI ($\lambda\lambda 5167, 5173, 5184$ \AA) and Fe I
($\lambda\lambda 5270, 5328$ \AA). The $0\farcs5$ wide and $5\farcm5$ long
Gemini slit is centered on NGC~3311, at
$\alpha=10\mbox{h}36\mbox{m}42.74\mbox{s}$ and
$\delta=-27\mbox{d}31\mbox{m}41.3\mbox{s}$ (J2000), along P.A. = $63^{\circ}$,
the direction of the galaxy major axis.  Its position is shown in
Fig.~\ref{geometry}.

\begin{figure}[hbt!] \centering
\includegraphics[width=7.0cm]{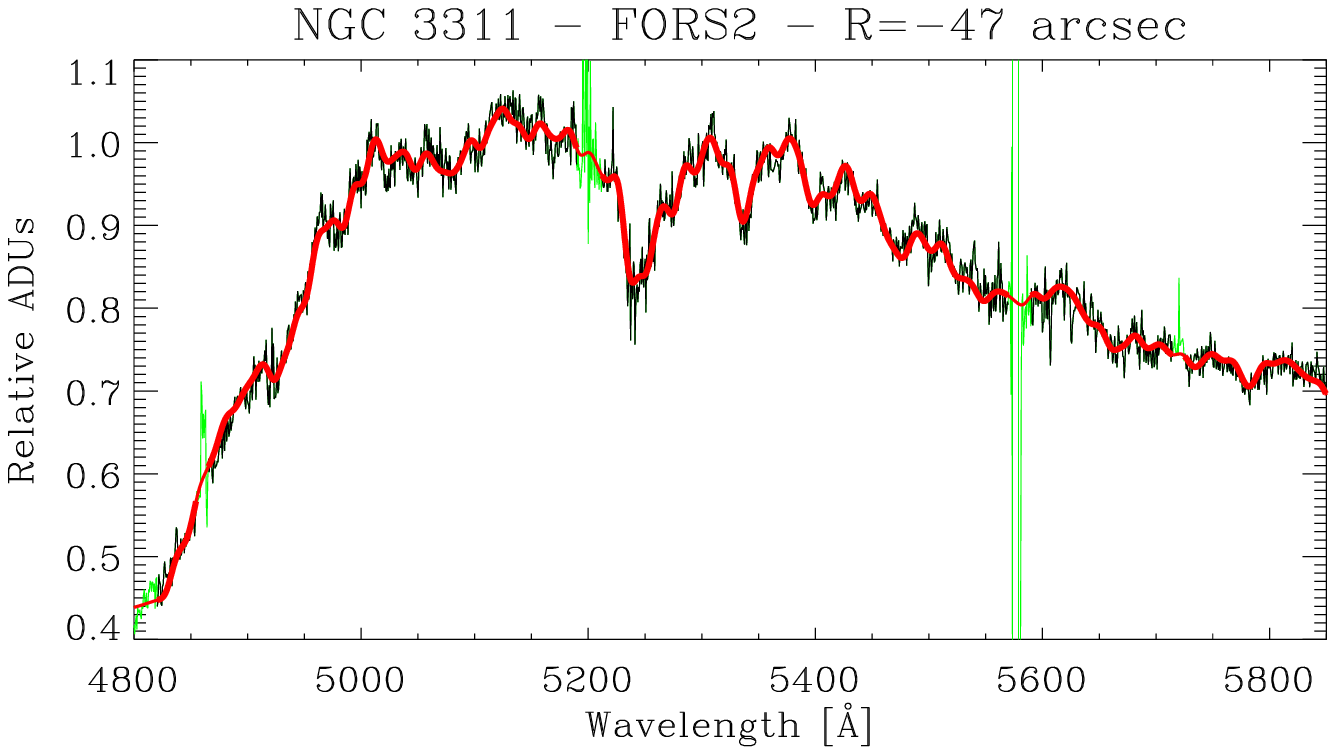}
\includegraphics[width=7.0cm]{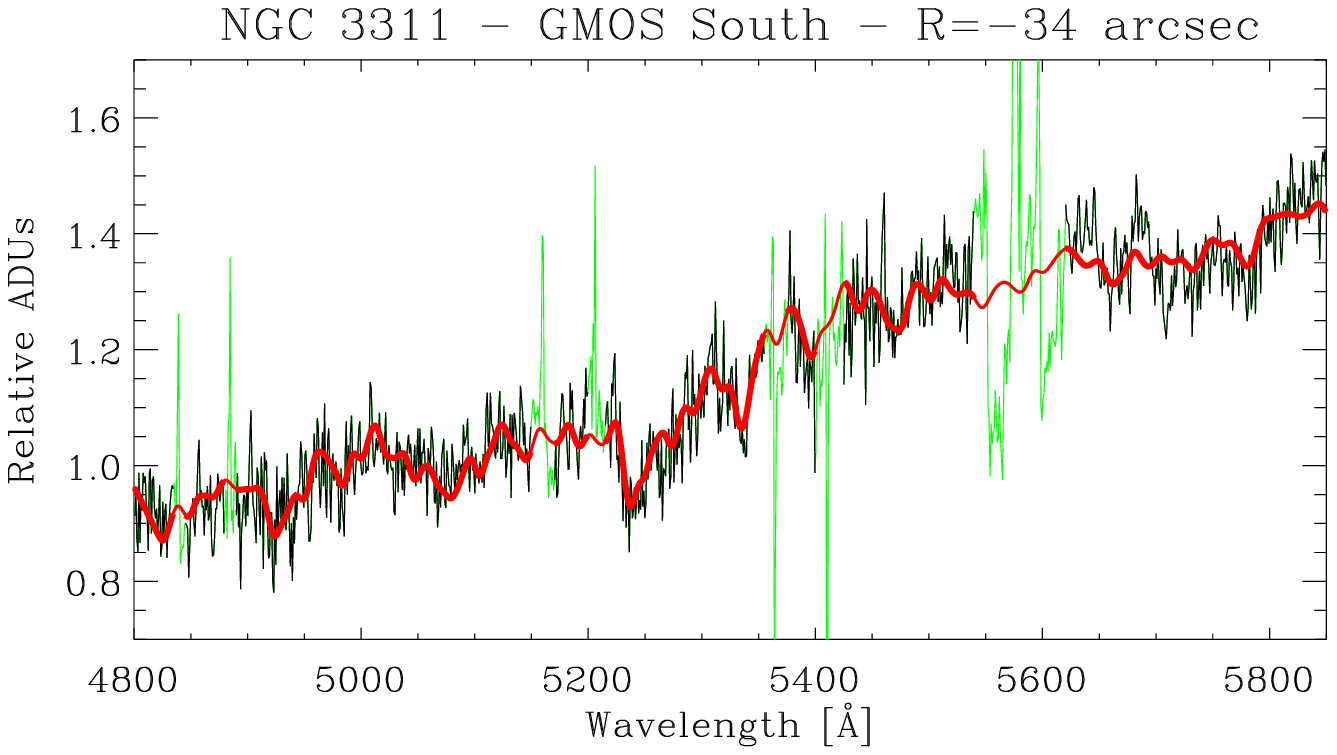}
\caption{Kinematic fits with PPXF of the spectra extracted at $-47''$ (VLT
FORS2) and at $-34''$ (Gemini GMOS). In {\it black} we display the galaxy
spectra, in {\it green} the wavelength range excluded from the fit because of
sky residuals, and in {\it red} the best-fit-broadened template model. All
spectra are normalized to the value of the best-fit model at 5100 \AA. }
\label{spectrafit}
\end{figure}

\section{Data reductions}\label{datared}

The data reduction of the FORS2 long slit spectra is carried out in $IRAF$.
After the standard operations of bias subtraction and flat-fielding, the spectra
are registered, co-added and wavelength calibrated. The edges of the FORS2 slit
reach well into sky regions, which are then used to interpolate the sky emission
in the regions covered by the stellar spectra.

In the low surface brightness regions, spectra are summed along the spatial
direction in order to produce one-dimensional spectra with an adequate S/N ratio
($\geq 20$ per \AA).  Four independent one-dimensional spectra are extracted
along the slit where the light is dominated by the halo of NGC~3311; of these,
two are from regions north and two from regions south of HCC~26, respectively.
We extract spectra from slit regions of $\sim31'' \times 1\farcs6$ and
$\sim25''\times 1\farcs 6$ at distances of about $54"$ and $47"$ from the center
of NGC~3311, and of $\sim58''\times 1\farcs6$ and $\sim36'' \times 1\farcs6$ at
central distances of about $100"$ and $67"$. We properly mask the spectra of
foreground stars in those areas.

The data reduction for the GMOS long slit spectra is carried out independently
here, also in $IRAF$ and with the standard tasks in the Gemini package. The
procedure is described in \cite{Loubser08} for the wavelength calibration and
background subtraction; also in this case the edges of the slit are used to
interpolate the sky emission in the regions covered by the stellar continuum.
Because our goal is to sample the kinematics well into the halo, the
one-dimensional spectra for the absorption line measurements are summed along
the slit direction so that a minimum S/N $\sim 20$ per \AA\ is obtained in each
radial bin, out to a radial distance of about $40"$ from the center of NGC~3311.

{\it Stellar kinematics}\label{ppxf} - The stellar kinematics is measured from
the extracted 1D spectra in the wavelength range 4800 \AA $< \lambda< 5800$ \AA
, using both the ``penalized pixel-fitting'' method (PPXF, \cite{Cappellari04})
and the Fourier correlation quotient (FCQ) method \citep{Bender90}, in order to
account for possible systematic errors and template mismatch.

\begin{figure}[hbt!] \centering
\includegraphics[width=9.0cm,clip=]{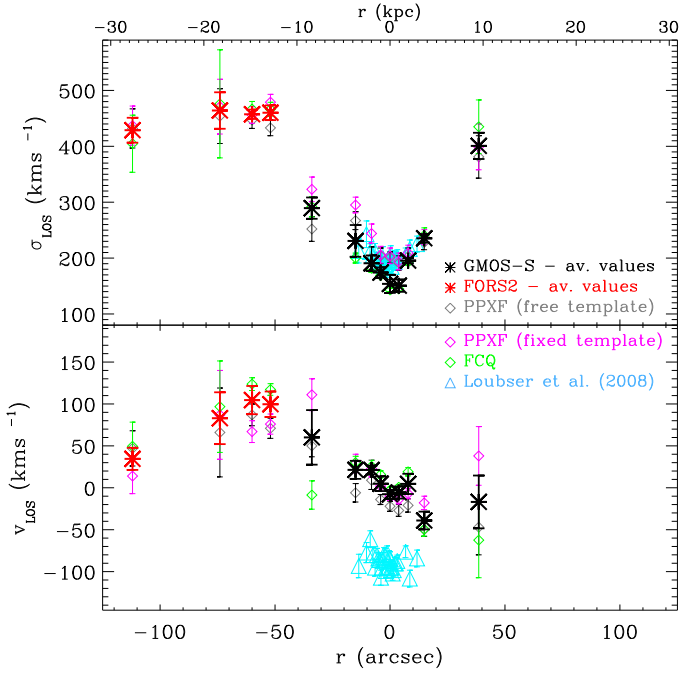}
\caption{Major axis line-of-sight velocity and velocity dispersion profiles for
NGC~3311 ($\mbox{P.A.} = 63^\circ$).  The open light blue triangles are the
values published by \cite{Loubser08}, based on Gemini-South GMOS data.  The
black asterisks are our independent measurements from these GMOS (archival)
spectra, and the red asterisks show measurements from the new VLT-FORS2 spectra.
These are weighted averages of three independent measurements which are obtained
with PPXF and FCQ as described in Sec.~\ref{datared} and shown separately as the
gray, magenta and green diamonds. The FORS2 data points are obtained from
averages over $\sim25''$ and $\sim31''$ in the inner regions and over $\sim36''$
and $\sim58''$ in the outer regions of the slit; see Fig.~\ref{geometry}.  These
off-axis measurements are plotted at their light-weighted average radii,
corrected for projection onto the major axis of NGC~3311 with an isophotal
flattening of 0.89.  Positive distances are south-west from the center of
NGC~3311 and negative values are north-east, along P.A.=$63^\circ$.}
\label{sigma3311}
\end{figure}

In the PPXF method, stellar template stars from the MILES library
\citep{Sanchez07} are combined to fit the one-dimensional extracted spectra; the
rotational velocity, the velocity dispersion and Gauss-Hermite moments (e.g. 
\cite{Gerhard93}) are measured simultaneously. Fig.~\ref{spectrafit} shows two
of the extracted spectra and the broadened templates fit by PPXF. In the FCQ
method, the rotational velocity and velocity dispersion are derived for each
extracted one-dimensional spectrum by assuming that the LOSVD is described by a
Gaussian plus third- and fourth- order Gauss-Hermite functions. Before to the
fitting procedure the MILES template spectra are smoothed to the GMOS and FORS2
spectral resolution with the measured broadening offsets. While FCQ provides
error estimates along with the measured kinematics, errors for the PPXF
kinematic parameters are calculated with a series of Monte Carlo simulations
adopting the appropriate S/N for each bin.

Because the stellar populations in cD halos may have different metal abundances
and ages from those of the inner regions \citep{Coccato10a,Coccato10b},
systematic effects caused by template mismatch must be evaluated and accounted
for. We therefore extract kinematic measurements with PPFX and FCQ as follows:
\begin{enumerate}
\item fit with PPFX the best stellar template from the MILES
  library in the central regions with the highest S/N, and extract $v$
  and $\sigma$ at all radii, using the same stellar template;
\item simultaneously fit the best stellar template,
  $v$ and $\sigma$ in each radial bin with PPFX;
\item adopt the respective best-fit PPXF stellar template
  to derive the LOSVD with FCQ for all radial bins;
\item finally, average rotational velocities $v$ and velocity
  dispersions $\sigma$ are computed as weighted means of the three
  values extracted in each radial bin as detailed above. Errors for
  these weighted average values are computed from those of the three
  measurements, but if the reduced
    $\chi^2=\frac{1}{(n-1)} \sum_{i=1}^{n}\frac{(x_i-\bar{x})^2}{
      \epsilon^2_i}$ is greater than one, they are increased by
    $\sqrt{\chi^2}$ in order to take into account systematic
    differences.  I.e., $\epsilon^2_{\bar{x}}=\frac{1}{\sum_{i=1}^{n}
      1/\epsilon^2_i}\times\chi^2$ where $\epsilon_i$,
    $\epsilon_{\bar{x}}$ are the errors on the individual measurements
    $x_i$ and the weighted mean ${\bar{x}}$, respectively.

\end{enumerate}

Mean velocities and velocity dispersions in all radial bins are listed in
Table~\ref{tablespectra}, and the profiles are shown in Fig.~\ref{sigma3311}
together with the previous measurements from \citet{Loubser08}. Table 1, which
is available in electronic form, contains the following information: source of
data (Col. 1), distance from galaxy center (Col. 2), P.A. (Col. 3), $v$,$\sigma$
with errors for each of the procedures 1.-4. described in the text, in Col.
(4-5), (6-7), (8-9), and (10-11), respectively. Heliocentric and relativistic
corrections have been applied to the mean velocities. The systemic velocity is
$3800\,\kms$ and has been subtracted.

In the central region of NGC~3311, our new velocity dispersion profile
marginally agrees with that of \citet{Loubser08}.  The new mean line-of-sight
velocity measurements agree with the systemic velocity of NGC~3311 obtained by
\citet{Misgeld08}, but have a systematic offset from the $v$ data of
\cite{Loubser08}, by about 91 km s$^{-1}$. The agreement between the new FORS2
measurements at $-47"$ and the revised value at $-34"$ from archive GMOS data
gives us confidence that the systematic effects from wavelength calibration
offsets, template mismatch, etc., are sufficiently small in the new, independent
data reductions. However, several tests have convinced us that the data do not
allow us to reliably determine full line-of-sight distributions (e.g., $h_3$,
$h_4$), which could be used to test for subcomponents, which one would expect in
particular at radii $\sim 30"-40"$.

\section{The kinematics of the NGC~3311 stellar halo}\label{kinN3311}

The combined new velocity dispersion profile for NGC~3311 reaches to $R_{mj}=
39''\simeq10\;\mbox{kpc}$ from the center of NGC~3311 along the galaxy's major
axis (P.A.=$63^\circ$), and to an off-axis distance of
$R=100''\simeq25\;\mbox{kpc}$ along the FORS2 slit. It shows a very unusual
steep rise with increasing radial distance from the galaxy center: from a
central value $\sigma_0 = 150\,\mbox{km s}^{-1}$, to $\sigma = 231\;\mbox{km
s}^{-1}$ at $R=15''\simeq3.7\;\mbox{kpc}$, and then on to a flat $\sigma_{out}
\simeq 450 \,\mbox{km s}^{-1}$ outside $R = 47'' = 12\;\mbox{kpc}$.  The steep
outward gradient is supported by two independent data sets and data reductions. 
The measurements of \cite{Loubser08} near the galaxy center had already hinted
at a positive gradient from $190\;\mbox{km s}^{-1}$ at $R=5''$ to
$\sim240\;\mbox{km s}^{-1}$ at a radius of $R=10''$, and data shown in Fig.~1 of
\citet{Hau+04} reach $\simeq 300\,\kms$ at $\sim25"$. With the new data we now
have very clear evidence of a dynamically hot stellar halo in NGC~3311.

\begin{figure}[hbt!] \centering
\includegraphics[width=7.0cm]{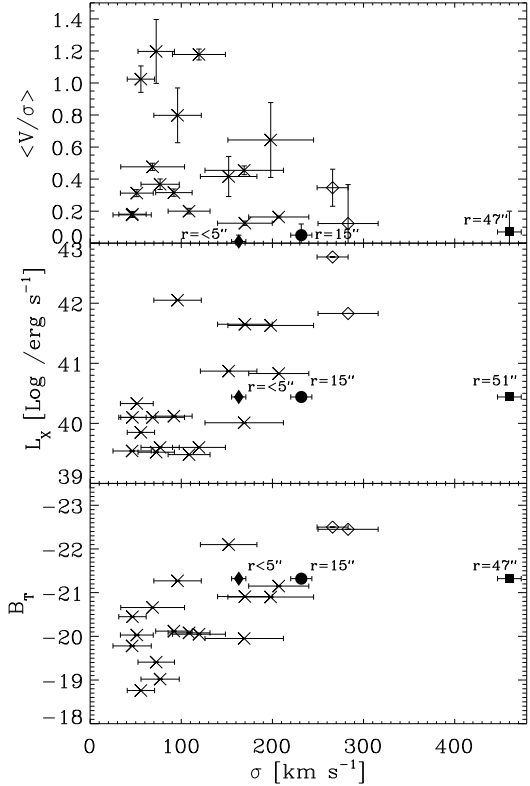}
\caption{Properties of the stellar halo of NGC~3311 compared with other
early-type galaxy halos: mean $<V/\sigma>$ ({\it upper panel}), total X-ray
luminosity ({\it central panel}), and B-band total magnitude ({lower panel})
against stellar velocity dispersion $\sigma$. Solid diamond, circle, and square
show the measured $\sigma$ of NGC~3311 at the center, $15"$ ($\simeq
3.7\;\mbox{kpc}$), and $47"$ ($\simeq12\;\mbox{kpc}$). Crosses show outermost
velocity dispersions from \cite{Coccato09}, and open diamonds for NGC~4889/4874
from \cite{Coccato10a}.}
\label{galvscluster}
\end{figure}

To put the extremely rapid rise of the velocity dispersion profile of NGC~3311
in context, we compare its kinematic properties with those of early-type galaxy
(ETG) halos mapped using planetary nebula data \citep{Coccato09} and with the
halos of the two Coma BCG galaxies NGC~4889, NGC~4874 from deep absorption line
spectroscopy \citep{Coccato10a}.  Fig.~\ref{galvscluster} shows the mean
$<V/\sigma>$, X-ray luminosity, and total absolute B-band magnitude for these
galaxies versus their outermost halo velocity dispersion.  For NGC~3311, we use
a bolometric X-ray luminosity within $50''\simeq 12\;\mbox{kpc}$, $L_X = 2.75
\times 10^{40}$ erg s$^{-1}$ (based on the flux in the 0.5-4.5 keV energy range
from \cite{Yamasaki02} and corrected to bolometric $L_X$ according to Table~1 of
\cite{Sullivan01}), and the total extinction corrected B-band magnitude (12.22)
from \cite{deVaucouleurs91}. For the velocity dispersion of NGC~3311, we use the
values at the center, at $15"$ ($\simeq 3.7\;\mbox{kpc}$) and at $47''$ ($\simeq
12\;\mbox{kpc}$).  Only the central $\sigma_0$ puts NGC~3311 in the middle of
the ETG distribution; $\sigma(47")$ deviates strongly, with a much larger
$\sigma$ than expected for the given $L_X$, $B_T$.

The natural interpretation for these results is that the outer stellar halo of
NGC~3311 is dominated by the central intracluster star component of the Hydra
cluster. This is supported by several pieces of evidence: (i) The steep rise of
the $\sigma$-profile; more isolated ETGs all have slightly or even steep falling
$\sigma$-profiles \citep{Coccato09}.  (ii) The saturation of $\sigma$ at $\simeq
12\;\mbox{kpc}$, outside of which the dynamically hot component dominates
completely. $\sigma(47")$ is $\sim60\%$ of the galaxy velocity dispersion in the
cluster core. (iii) The large amount of dark matter inferred from X-ray
observations around NGC~3311 ($\sim 10^{12} \msun$ within 20 kpc,
\cite{Hayakawa04}).

In recent cosmological hydrodynamic simulations of cluster formation,
\cite{Dolag10} applied a kinematic decomposition to the stellar particles around
cD galaxies. With a double Maxwellian fit to the velocity histogram of star
particles centered on a simulated cD, they were able to separate an inner,
colder Maxwellian distribution associated with the central galaxy, and an outer,
hotter component of stars that orbit in the cluster potential. For both
components they derived radial density profiles and, fitting Sersic profiles,
found that the inner stellar component is much steeper than the outer diffuse
stellar component. A comparison with these simulations indicates that the steep
velocity dispersion gradient in the halo of NGC~3311 traces the transition from
central galaxy stars to the diffuse intracluster stellar component. In the
NGC~3311 halo, the transition between the two occurs at smaller radii than in
other BCGs in nearby clusters, in the range between 4 and 12 kpc.

NGC~3311 appears to have a similar halo as the cD galaxy NGC~6166 in the
Abell~2199 cluster \citep{kelson02}, whose $\sigma$-profile rises to cluster
values at $R\sim 60$ kpc. But NGC~3311 is even more extreme; it is a fairly
small galaxy, based on its central $\sigma_0=150$ $\mbox{km s}^{-1}$, and it is
already dominated by the surrounding cluster component at $R\sim 12$ kpc. 
Presumably, this is because the core of the ``relaxed'' Hydra cluster has had
time to collapse onto the galaxy. For comparison, the two BCG galaxies in the
Coma cluster core, which have a nearly constant $\sigma$-profile
\citep{Coccato10a}, may be in the middle of an ongoing merger \citep{Gerhard07},
so that their previous subcluster halos would have been stripped and a new
cluster halo could have been built only after the merger was completed; and in
the outer halo of the more isolated M87, the velocity dispersion appears to drop
\citep{Doherty09} towards the edge.

\section{Conclusions} \label{Conclu}

Based on two independent long-slit data sets and reductions, we find a steep
gradient in the velocity dispersion profile of the central galaxy NGC~3311 in
the Hydra I cluster, from $\sigma_0\simeq 150$ $\mbox{km s}^{-1}$ to
$\sigma_{out} \simeq 450$ km s$^{-1}$ outside 12 kpc (60\% of the velocity
dispersion of the galaxies in the surrounding cluster).

The new data provide evidence that NGC~3311 is a fairly small galaxy dominated
by a large envelope of intracluster stars already beyond $R\sim12$ kpc, whose
orbits are dominated by the cluster dark matter potential.  Comparison with
other BCG galaxies shows a wide range of dynamical behavior in their halos.

\begin{acknowledgements}
  The authors wish to thank the ESO VLT staff for their support during the
observations and the referee for a constructive report.  This research has made
use of the Gemini archive data, the NASA/IPAC Extragalactic Database (NED),
which is operated by the Jet Propulsion Laboratory, California Institute of
Technology, under contract with the National Aeronautics and Space
Administration.
\end{acknowledgements}

\bibliography{gventimi}

\newpage

\def\kms{$\rm km\;s^{-1}$}
\begin{landscape}
\begin{table}
\caption{Measured mean velocities and velocity dispersions for NGC~3311. For
details see text. The galaxy's systemic velocity $V_{sys} = 3800$ \kms\ has been
obtained by a linear fit to the velocities in the central $20''$ and has then
been subtracted from the measurements. This value includes heliocentric and
relativistic corrections.}
\centering
\begin{tabular}{l c c c c c c c c c c}
\hline
\noalign{\smallskip} 
Instr. & R          &     P.A.   &$V_{\rm ppxf 1}$   & $\sigma_{\rm ppxf 1}  $     &$V_{\rm ppxf 2} $  & $\sigma_{\rm ppxf 2}$     &$V_{\rm FCQ}$  & $\sigma_{\rm FCQ}$    & $<V>$     &  $<\sigma>$               \\
       & (arcsec)   &            &  (\kms)           &  (\kms)                   &(\kms)            & (\kms)                  &(\kms)        & (\kms)                 & (\kms)     &   (\kms)                 \\
(1) & (2) & (3) & (4) & (5) & (6) & (7) & (8) & (9) & (10) & (11)\\
\hline                                                                                                                                                                                                  
\noalign{\smallskip}                                                                                                                                                                                    
GMOS-S & $  38.6  $ &$ 63^\circ$ &  $ 38 \pm 33 $  & $399 \pm 38 $               &$-47 \pm 33 $    & $381 \pm 38  $            & $-62 \pm 45 $ &  $435 \pm 48 $       &$ -17 \pm31 $&$ 401 \pm 23 $         \\
GMOS-S & $  14.89 $ &$ 63^\circ$ &  $-18 \pm  8 $  & $238 \pm 13 $               &$-50 \pm  8 $    & $228 \pm 13  $            & $-49 \pm  8 $ &  $241 \pm 14 $       &$ -39 \pm11 $&$ 235 \pm  8 $         \\
GMOS-S & $   7.88 $ &$ 63^\circ$ &  $ -7 \pm  8 $  & $209 \pm 15 $               &$-21 \pm  8 $    & $203 \pm 15  $            & $ 19 \pm  5 $ &  $193 \pm  5 $       &$   5 \pm12 $&$ 195 \pm  5 $         \\
GMOS-S & $   3.8  $ &$ 63^\circ$ &  $-14 \pm  7 $  & $194 \pm 13 $               &$-27 \pm  7 $    & $191 \pm 13  $            & $  1 \pm  3 $ &  $145 \pm  3 $       &$  -5 \pm7  $&$ 151 \pm 11 $        \\
GMOS-S & $   0    $ &$ 63^\circ$ &  $-10 \pm  6 $  & $204 \pm 10 $               &$-22 \pm  6 $    & $201 \pm 10  $            & $ -3 \pm  3 $ &  $143 \pm  3 $       &$  -8 \pm5  $&$ 154 \pm 16 $        \\
GMOS-S & $  -4.09 $ &$ 63^\circ$ &  $  2 \pm  6 $  & $206 \pm 14 $               &$-14 \pm  6 $    & $204 \pm 14  $            & $ 16 \pm  4 $ &  $169 \pm  4 $       &$   5 \pm9  $&$ 174 \pm  9 $        \\
GMOS-S & $  -8.05 $ &$ 63^\circ$ &  $ 23 \pm 11 $  & $244 \pm 20 $               &$  9 \pm 11 $    & $200 \pm 20  $            & $ 26 \pm  7 $ &  $182 \pm  8 $       &$  21 \pm5  $&$ 191 \pm 14 $        \\
GMOS-S & $  -15   $ &$ 63^\circ$ &  $ 29 \pm 11 $  & $295 \pm 16 $               &$ -6 \pm 11 $    & $267 \pm 16  $            & $ 30 \pm  7 $ &  $200 \pm  9 $       &$  22 \pm11 $&$ 231 \pm 28 $        \\
GMOS-S & $  -34   $ &$ 63^\circ$ &  $111 \pm 13 $  & $323 \pm 22 $               &$ 50 \pm 13 $    & $252 \pm 22  $            & $ -8 \pm 17 $ &  $292 \pm 18 $       &$  60 \pm33 $&$ 289 \pm 19 $        \\
FORS2  & $  -47   $ &$ 45^\circ$ &  $ 76 \pm 12 $  & $479 \pm 14 $               &$ 71 \pm 12 $    & $433 \pm 14  $            & $117 \pm  7 $ &  $466 \pm 12 $       &$ 100 \pm15 $&$ 460 \pm 13 $        \\
FORS2  & $  -54   $ &$ 83^\circ$ &  $ 67 \pm 13 $  & $456 \pm 15 $               &$ 87 \pm 13 $    & $447 \pm 15  $            & $124 \pm  8 $ &  $467 \pm 14 $       &$ 105 \pm17 $&$ 457 \pm  8 $        \\
FORS2  & $  -67   $ &$  6^\circ$ &  $ 87 \pm 53 $  & $471 \pm 49 $               &$ 66 \pm 53 $    & $454 \pm 49  $            & $ 97 \pm 55 $ &  $476 \pm 97 $       &$  83 \pm31 $&$ 464 \pm 33 $        \\
FORS2  & $ -100   $ &$114^\circ$ &  $ 14 \pm 21 $  & $437 \pm 35 $               &$ 47 \pm 21 $    & $432 \pm 35  $            & $ 50 \pm 29 $ &  $405 \pm 51 $       &$  35 \pm13 $&$ 429 \pm 22 $        \\
\hline
\end{tabular}
\label{tablespectra}
\begin{minipage}{24cm}
Notes --  Col. 1: instrument. 
          Col. 2: Radial distance from center of NGC 3311. 
          Col. 3: Position angle of data with respect to NGC 3311's center. 
          Col. 4: Velocity measured with PPXF (using template determined at $R=0"$), relative to the galaxy systemic velocity.
          Col. 5: Velocity dispersion measured with PPXF (using template determined at $R=0"$).
          Col. 6: Velocity measured with PPXF (free template), relative to the galaxy systemic velocity.
          Col. 7: Velocity dispersion measured with PPXF (free template).
          Col. 8: Velocity measured with FCQ, relative to the galaxy systemic velocity.
          Col. 9: Velocity dispersion measured with FCQ.
          Col. 10: Weighted average velocity.
          Col. 11: Weighted average velocity dispersion.
\end{minipage}
\end{table}
\end{landscape}

\end{document}